\documentclass[preprint,showpacs,preprintnumbers,superscriptaddress]{revtex4-1}
\usepackage{amsmath,amssymb,graphicx,epstopdf,bm,subfigure,float,siunitx,color}
\usepackage[figuresright]{rotating}
\usepackage{mathrsfs}

\newcommand{\be}{\begin{equation}}
\newcommand{\ee}{\end{equation}}
\newcommand{\1}{\left}
\newcommand{\2}{\right}
\def\({\left(}
\def\){\right)}
\def\[{\left[}
\def\]{\right]}

\newcommand{\dif}{\,\mathrm{d}}

\newcommand{\p}{\partial}

\newcommand{\m}{\mu}
\newcommand{\n}{\nu}
\newcommand{\al}{\alpha}

\renewcommand{\th}{\theta}

\newcommand{\na}{\nabla}

\begin{document}
\title{Damping of gravitational wave in $f(R)$ gravity }

\author{Haiyuan Feng\footnote{Corresponding author} }

\email{Email address: fenghaiyuanphysics@gmail.com}
\affiliation{Department of Physics, Southern University of Science and Technology, Shenzhen 518055, Guangdong, China}

\author{Laiyuan Su}
\email{Email address: 12131268@mail.sustech.edu.cn}
\affiliation{Department of Physics, Southern University of Science and Technology, Shenzhen 518055, Guangdong, China}

\author{Rong-Jia Yang\footnote{Corresponding author}}
\email{Email address: yangrongjia@tsinghua.org.cn}
\affiliation{College of Physical Science and Technology, Hebei University, Baoding 071002, China}

\author{Wei-Qiang Chen\footnote{Corresponding author}}
\email{Email address: chenwq@sustech.edu.cn}
\affiliation{Department of Physics, Southern University of Science and Technology, Shenzhen 518055, Guangdong, China}

\begin{abstract}
 We investigate the damping of  gravitational waves (GW) in $f(R)$ gravity by matter. By applying the kinetic theory, we examine the first-order approximation of the relativistic Boltzmann equation.
In the flat spacetime, 
we derive the evolution equations for waves in $f(R)$ gravity and demonstrate that Landau damping is absent while collision damping is present.
In the Friedmann-Robertson-Walker (FRW) cosmology, we also examine the dynamical equations for the two modes. Furthermore, in the model $f(R) = R + \al R^2$, we investigate the effect of the mass term on wave amplitude decay within the neutrino system. We observe that the tensor mode with $m = 1 \, \text{eV}$ exhibits faster decay compared to other cases, while the scalar mode with $m = 1 \, \text{eV}$ appears to suppress decay.


\end{abstract}

\maketitle

\section{Introduction}
The detection of gravitational waves (GW) in the universe has significantly advanced the development of modern astronomy and physics. Continuous observations provide crucial data to restrict characteristics of astrophysical sources \cite{Bird_2016,Woosley:2016nnw,Loeb:2016fzn,Li:2016iww,Zhang:2016rli,Yamazaki:2016fyr,Perna:2016jqh,Morsony:2016upv,Liu:2016olx}, as well as to test general relativity (GR) \cite{LIGOScientific:2016lio,Blas:2016qmn,Ellis:2016rrr,Wu:2016igi,Collett:2016dey,Lombriser:2015sxa}.
The interaction of GW with matter,  though often neglected, has been investigated throughout the history. Hawking first calculated the damping rate of GW as  $\gamma=16\pi G\eta$ by viewing matter as a fluid, with the viscosity $\eta$  \cite{Hawking:1966qi}. Subsequently, Ehlers et.al proved  that GW traveling through perfect fluid didn't suffer from dispersion or dissipation \cite{Svitek:2008pd}. In the collisionless limit, the damping rate of GW by non-relativistic particles was shown to be related to the velocity and the number density \cite{osti_5063362}. By linearizing the Boltzmann equation and accounting for the collision term,  a unified treatment for damping from collision and the Landau damping was provided by Ref.\cite{Baym:2017xvh}. Landau damping, initially introduced to investigate the dispersion relation in the plasma systems \cite{Landau:1956zuh,Tanaka:1993mw,Sun:2019qno}, and then generalized to the research of large-scale galaxy clusters \cite{Eggen:1962dj,Ostriker:1966zz}, was shown to be vanish for GW in flat spacetime.  It is a well-established result that transverse GW are not absorbed by non-collisional massive media \cite{PhysRevD.97.123506,Gayer:1979ff,Asseo:1976bc,PhysRevD.7.2863}. Lambda damping occurs only in the presence of viscosity \cite{Anile1978HighfrequencyGW,Hawking:1966qi,Madore:1973xy}, or when a medium consisting of massless particles is considered \cite{Chesters:1973wan,Stefanek:2012hj}. Though this effect holds considerable conceptual importance, it is practically minimal and restricted to particular wavelengths. A more intriguing approach is to explore the various modified gravity that enable the emergence of an additional massive scalar modes. In this context, together with transverse polarization, scalar modes are responsible for a longitudinal polarization, which we anticipate could interact with particles in the medium. To investigate whether this polarization may arise Landau damping, we will employ kinetic theory approach to analyze the interaction between scalar modes and collisionless particle distribution.

In addition, GW significantly influenced the development of the early universe. The observation of cosmic tensor fluctuation by measurements of microwave background polarization is the most precise method to for testing the inflationary universe.  Initially, Weinberg outlined the primary approach for calculating the influence of collisionless, massless neutrinos during the radiation-dominated era \cite{Weinberg:2003ur}. The literature demonstrates that the damping effect of free-streaming neutrinos on the GW spectrum can be quite significant, with up to a 35.6$\%$ reduction in amplitude
\cite{bashinsky2005coupledevolutionprimordialgravity,PhysRevD.72.088302,PhysRevD.77.063504,PhysRevD.73.123515,PhysRevD.75.104009,PhysRevD.86.123502}. Based on Weinberg's conclusions, Ref.\cite{Stefanek:2012hj} developed a set of analytical solutions using modal expansions with spherical Bessel functions as bases, providing a robust framework for further investigations. Additionally, the damping effect of GWs in cold dark matter has been explored by Ref.\cite{Flauger:2017ged}, which also included considerations for mass-relativistic particles. This research highlighted the complexities involved in the interplay between dark matter and GW, underscoring the importance of accounting for various particle masses in these calculations. Moreover, investigating the damping of GW in modified gravity, particularly those involving additional scalar modes, has also become increasingly significant. Therefore, we aim to determine whether the damping affects the GW in modified gravity.



To address the cosmological constant problem, the $f(R)$ theory was proposed. This model has two significant advantages: the action is sufficiently general to encompass some of the fundamental properties of higher-order gravity while remaining simple enough to be easily handled; it appears to be capable of averting the long-known and catastrophic Ostrogradski instability \cite{Woodard:2006nt}. Especially, choosing the simplest $f(R)=R+\al R^2$ $(\al>0$), can explain the universe's accelerated expansion and serves as a candidate for an inflationary field \cite{STAROBINSKY198099}.  In conclusion, $f(R)$ theory is a significant theoretical framework for modifying gravity and is worth further investigation.

In this paper, we investigate the damping of $f(R)$ GW in the presence of matter and determine the dispersion relation by considering contributions from the collision term. In Section II, we introduce linearized $f(R)$ theory and provide wave equations for tensor and scalar modes in Minkowski spacetime. In Section III, we apply kinetic theory to obtain the first-order approximation of the relativistic Boltzmann equation. We calculate the anisotropic part of the spatial component of the energy-momentum tensor and derive the dispersion relations of the two modes using the relaxation time approximation. Additionally, we derive damping coefficients in the collision-dominant case and investigate Landau damping. In Section IV, within the context of FRW cosmology, we establish the damping equations for two modes and demonstrate that Landau damping exists in the FRW background.  In Section V, we numerically investigate the effect of neutrino mass on the damping of the two modes within the specific model  $f(R) = R + \al R^2$. Finally, in Section VI, we present concluding remarks on our findings. Throughout the article, we use the signature convention $(-,+,+,+)$ for the spacetime metric. Spacetime dimensions are labeled with Greek indices, $\m$=0, 1, 2, 3; spatial dimensions are labeled with Latin indices $i=1, 2, 3.$


\section{Linearized gravitational waves in $f(R)$ theory}
The action of $f(R)$ theory has the following form
\be
\label{1}
S[g_{\mu\nu}]=\frac{1}{2\kappa^2}\int d^{4}x\sqrt{-g}f(R)+\int d^{4}x\sqrt{-g}L_m,
\ee
with $\kappa^2=8\pi G$, $L_m$ is the Lagrangian of matter. 
The field equation can be obtained by varying the above action \cite{Kalita:2021zjg},
\be
\label{2}
F(R)R_{\m\n}-\frac{1}{2} f\(R\)g_{\m\n}-\na_{\m}\na_{\n}F(R)+g_{\m\n}\Box F(R)=\kappa^2 T^{(M)}_{\m\n},
\ee
where $F(R)=\frac{\dif f(R)}{\dif R}$, and $\Box$ is the d'Alembertian operator. The energy-momentum tensor $T^{(M)}_{\m\n}\equiv -\frac{2}{\sqrt{-g}}\frac{\delta L_m}{\delta g^{\m\n}}$  satisfies the continuity equation $\na_{\m}T^{(M)\m\n}=0$. We rearrange the preceding equations to get
\be
\1\{\begin{split}
\label{3}
&G_{\m\n}=\kappa^2\(T^{(eff)}_{\m\n}+T^{(M)}_{\m\n}\),\\
&\kappa^2T^{(eff)}_{\m\n}\equiv \frac{g_{\m\n}}{2}\(f(R)-R\)+\na_{\m}\na_{\n}F(R)-g_{\m\n}\Box F(R)+\(1-F(R)\)R_{\m\n},
\end{split}\2.
\ee
with Einstein tensor $G_{\m\n}=R_{\m\n}-\frac{1}{2}g_{\m\n}R$, and fulfills the Bianchi identity $\na_{\m}G^{\m\n}=0$. It can be proved that the contribution of curvature $T^{(eff)}_{\m\n}$ also obey $\na_{\m}T^{(eff)\m\n}=0$. It is possible to get the trace of Eq. \eqref{3} as
\be
\label{4}
3\Box F(R)+F(R)R-2f(R)=\kappa^2T^{(M)},
\ee
which clearly demonstrates the difference from Einstein's trace equstion $R=-\kappa^2 T$. The presence of the term $\Box F(R)$ leads to additional degrees of freedom in the propagation.

To investigate the equation of $f(R)$ GW, the metric $g_{\m\n}$ and Riemann curvature scalar $R$ in Minkowski spacetime are perturbed as follows
\be
\1\{\begin{split}
\label{5}
&g_{\m\n}=\eta_{\m\n}+h_{\m\n},\\
&R=R_0+\delta R,
\end{split}\2.
\ee
where tensor perturbation $h_{\m\n}$ is restricted by $|h_{\m\n}|\ll|\eta_{\m\n}|$. The background curvature and scalar perturbation are denoted by $R_0$ and $\delta R$, respectively. As can be shown, different from GW in GR, the perturbation has the form $h_{\m\n}=\bar{h}^{TT}_{\m\n}+h^{S}_{\m\n}$, where $\bar{h}^{TT}_{\m\n}$ represents the transverse-traceless (TT) part of the perturbation. It satisfies $\partial_{i}\bar{h}^{ijTT}=0$, $\bar{h}^{iTT}_{i}=0$, and $h^{S}_{\m\n}=-\phi \eta_{\m\n}$. $\phi\equiv \frac{F'(R_0)\delta R}{F(R_0)}$ represents the scalar degree of freedom \cite{Kalita:2021zjg}. This mode can manifest as a "breathing mode," characterized by isotropic spatial expansion and contraction. It also represents an additional polarization state in GW, distinct from the transverse modes ($+$ and $\times$ polarizations) in GR, thereby serving as a key signature of $f(R)$ gravity. Meanwhile, the linearized field equations are given by \cite{CAPOZZIELLO2008255,PhysRevD.95.104034,RevModPhys.82.451,doi:10.1142/S0218271814500370,PhysRevD.99.104046}
\be
\1\{\begin{split}
\label{6}
&\Box\bar{h}^{TT}_{ij}=-2\kappa^2\Pi^{(1)}_{ij},\\
&\Box\phi-M^2\phi=\frac{\kappa^2}{3F(0)}T^{(1)},
\end{split}\2.
\ee
where
\be
\label{7}
M^2\equiv\frac{F(0)}{3F'(0)},
\ee
 is the square of the effective mass,
 and $\Pi^{(1)}_{ij}$ is the linear part of the anisotropic part of the spatial components of energy-momentum tensor $T_{ij}$ $\(T^{i}_{j}=\Pi^{i}_{j}+\frac{1}{3}\delta^{i}_{j}\sum^{3}_{k=1}{T^{k}_{k}}\)$. It couples with GW and satisfies $\Pi^{i}_{i}=0$, $\partial_{i}\Pi^{ij}=0$. It is obvious from Eq.\eqref{6} that when the effective mass $M$ approaches infinity, the system no longer has the excitation of the scalar mode and returns to the tensor mode of GR.  Consequently, the number of polarizations in $f(R)$ gravity is three \cite{universe4080085,PhysRevD.93.124071}.

\section{Landau and collision-dominated damping of  $f(R)$ gravitational wave  in Flat spacetime}
The relativistic Boltzmann equation describes the time evolution of the distribution function in a system of relativistic particles and is widely used in cosmology, plasma physics, and high-energy astrophysics. Specially, the properties of on-shell particles vary depending on the specific physical scenario \cite{Jeon_1996,Blaizot_2002,Kapusta:2006pm,Mathieu_2014,Stockamp:2004qu,Epelbaum:2015vxa,kremer2014theoryapplicationsrelativisticboltzmann,Bernstein:1988bw,Dodelson:2003ft,Ma_1995,Liebendoerfer:2003es,Janka:2006fh}. For instance, in neutral gas, the particles are atoms and molecules, and their evolution are primarily driven by collisions. In plasma, interactions occur through the electromagnetic field generated by charged particles. In astrophysics, particles constitute stars, galaxies, and even clusters of galaxies, and their interactions are governed by gravity.

To calculate the anisotropic stress $\Pi_{ij}$ and the energy-momentum tensor trace $T$, we consider the relativistic Boltzmann equation \cite{Jeon:1995zm,Blaizot:2001nr,Kapusta:2006pm,2014arXiv1404.7083K,Dodelson:2003ft}
\be
\label{8}
p^{\m}\frac{\p f}{\p x^{\mu}}-g_{ij}\Gamma^{i}_{\m\n}p^{\m}p^{\n}\frac{\p f}{\p p_j}=C\[f\],
\ee
where distribution function $f(x^i, p_j)$ describes the probability of the spatial distribution. $\Gamma^{i}_{\m\n}$ is the connection coefficient and $p^{\m}$ represents the four-momentum of a single particle with on-shell condition $g^{\m\n}p_{\m}p_{\n}=-m^2$.
$C\[f\]$ is the collision term which represents the instantaneous change in the distribution function due to close-range collisions. 
To simplify the structure of the collision term while preserving its fundamental characteristics, Anderson and Witting (AW) model has been proposed \cite{ANDERSON1974466}. This model is derived using the relaxation time approximation (or Bhatnagar-Gross-Krook approximation) \cite{PhysRev.94.511}. A comparison of the Navier-Stokes transport coefficients calculated from the AW model with those obtained from the full Boltzmann equation suggests that the values of these coefficients will not differ greatly from each other \cite{ANDERSON1974466,anderson1974relativistic,anderson1976relativistic}. The collision term is expressed as
\be
\label{9}
C\[f\]=-\frac{p^{\m}u_{\m}}{\tau_c}\(f_h-f\),
\ee
$\tau_c$ is the particle's average collision time, which depends on the average relative velocity between two particles and the collision cross-section, and $u^{\m}$ denotes the macroscopic fluid's four-velocity \cite{Romatschke:2015gic}.  Therefore, four-velocity could currently be written as $u^{\m}=\(1,0,0,0\)$ in the fluid's rest reference frame. The distribution function of the local equilibrium ($f_h$) is defined as
\be
\label{10}
f_{h}=\frac{g}{e^{\frac{-p^{\m}u_{\m}}{\mathscr{T}}}\pm 1},
\ee
where 
$\pm$ corresponds to fermions or bosons, $g$ is the number of degrees of freedom for the varieties of single particles, and $\mathscr{T}$ is the temperature. Using geodesic equation of particles, we can simplify Eq. \eqref{8} as
\be
\label{11}
\frac{\p f}{\p t}+\frac{p^m}{p^t}\frac{\p f}{\p x^m}+\frac{\dif p_m}{\dif t}\frac{\p f}{\p p_m}=\frac{1}{\tau_c}\(f_h-f\),
\ee
with
\be
\label{12}
\frac{\dif p_m}{\dif t}=\frac{1}{2}\frac{\p g_{\m\n}}{\p x^{m}}\frac{p^{\m}p^{\n}}{p^0}.
\ee

 Additionally, we will apply the dynamic perturbation approach to determine the formulation of the induced energy-momentum tensor. First, starting with $h_{\m\n}=\bar{h}^{TT}_{\m\n}-\phi\eta_{\m\n}$,
the perturbation on-shell condition can be expressed as
\be
\1\{\begin{split}
\label{13}
&\epsilon=\epsilon_0+\delta\epsilon,\\
&\epsilon_0\equiv p^0 =\sqrt{m^2+p^2},\\
&\delta\epsilon=\frac{\delta g^{\m\n}p_{\m}p_{\n}}{2\epsilon_0}=\frac{-\bar{h}^{ijTT}p_ip_j-m^2\phi}{2\epsilon_0}.
\end{split}\2.
\ee
Subsequently, we adopt the first-order perturbation $h^{\m\n} \eta_{\m\al} \eta_{\n\beta}=-h_{\al\beta}$. By substituting Eq. \eqref{13} into Eq. \eqref{12}, we derive
\be
\label{14}
\frac{\dif p_m}{\dif t}=\frac{1}{2p^0}\(p_k p_l\frac{\p \bar{h}^{klTT}}{\p x^m}+m^2\frac{\p\phi}{\p x^m}\).
\ee
Next, we handle the perturbed distribution function $f=f_0(p)+\delta f(x^i,p_j,t)$ according to Ref. \cite{Baym:2017xvh}. By ignoring all higher order terms, the linearized Boltzmann equation is
\be
\label{15}
\frac{\p \delta f}{\p t}+\frac{p^m}{p^0}\frac{\p \delta f}{\p x^m}+\frac{1}{2p^0}\(p_k p_l\frac{\p \bar{h}^{TT}_{kl}}{\p x^m}+m^2\frac{\p\phi}{\p x^m}\)\frac{\p f_0(p)}{\p p_m}=-\frac{1}{\tau_c}\(\delta f-\delta f_h\).
\ee

It is worth emphasizing that $\delta f_h$ represents the deviation of the distribution function from local equilibrium and the absence of perturbation. This deviation can be expanded into a first-order small quantity as $\delta f_h=\frac{\p f_0}{\p \epsilon}\delta\epsilon$ using Taylor's formula. By Fourier transforming $\bar{h}^{TT}_{ij}(\vec{r},t)= e^{i\vec{k}\cdot\vec{r}-i\omega t}\bar{h}^{TT}_{ij}(\omega,\vec{k})$ and $\phi(\vec{r},t)=e^{i\vec{k}\cdot\vec{r}-i\omega t}\phi(\omega,\vec{k})$, we obtain
\be
\label{16}
\delta f(\omega,\vec{k})=\frac{f'(p)}{2p} \frac{\bar{h}^{ijTT}p_ip_j\(-\frac{1}{\tau_c}-\frac{i \vec{p}\cdot\vec{k}}{p^0}\)-m^2\phi\( \frac{i\vec{p}\cdot\vec{k}}{p^0}+\frac{1}{\tau_c}\)}{\(-i\omega+\frac{i \vec{p}\cdot\vec{k}}{p^0}+\frac{1}{\tau_c}\)},
\ee
where $f'(p)$ denotes the derivation  with respect to $p$. Conclusively, since the induced anisotropic stress tensor is assessed in terms of the distribution function $f$, the dynamical system is comprehensively characterised by \cite{Baym:2017xvh,Flauger:2017ged,Hwang:2005hb}
\be
\1\{\begin{split}
\label{17}
&\Pi^{(1)}_{ij}=\int \frac{\dif^3p}{(2\pi)^3}\frac{p_ip_j}{\epsilon_0}\bar{\delta} f,\\
&T^{(1)}=-m^2\int \frac{\dif^3p}{(2\pi)^3}\frac{1}{\epsilon_0}\bar{\delta}f ,
\end{split}\2.
\ee
where $\bar{\delta}f\equiv\delta f-\delta f_h$ should be interpreted as the effect of the distribution function's own variation, since the total shift is the sum of the distribution function's own and the transformation  caused by $h_{ij}$.  
We can determine the expression by inserting Eq. \eqref{16} into Eq. \eqref{17}, which follows
\be
\label{18}
\Pi^{(1)}_{ij}=\bar{h}^{klTT}\int \frac{\dif^3p}{(2\pi)^3}\frac{p_kp_lp_i p_j f'_0(p)}{2p\epsilon_0}\(\frac{-i\omega}{-i\omega+\frac{i \vec{p}\cdot\vec{k}}{p^0}+\frac{1}{\tau_c}}\),
\ee
and
\be
\begin{split}
\label{19}
T^{(1)}&=-m^2\int \frac{\dif^3p}{(2\pi)^3}\frac{f'_0(p)}{2p\epsilon_0}\[\bar{h}^{klTT}p_kp_l\(\frac{-i\omega}{-i\omega+\frac{i \vec{p}\cdot\vec{k}}{p^0}+\frac{1}{\tau_c}}\)-m^2\phi\( \frac{i\omega}{-i\omega+\frac{i \vec{p}\cdot\vec{k}}{p^0}+\frac{1}{\tau_c}}\)\]\\
&=m^4\phi(\omega,\vec{k})\int \frac{\dif^3p}{(2\pi)^3}\frac{f'_0(p)}{2p\epsilon_0}\( \frac{i\omega}{-i\omega+\frac{i \vec{p}\cdot\vec{k}}{p^0}+\frac{1}{\tau_c}}\),
\end{split}
\ee
Based on the angular integration, the contribution of $\bar{h}^{klTT}$ in Eq. \eqref{19} is zero. The first term on the right side of Eq. \eqref{18} can be shown to be proportional to $\bar{h}^{TT}_{ij}$,
which follows
\be
\label{20}
\begin{split}
\Pi^{(1)}_{ij}=\bar{h}^{TT}_{ij}\int \frac{\dif^3p}{(2\pi)^3}\frac{(p_k p_l)^2 f'_0(p)}{p\epsilon_0}\(\frac{-i\omega}{-i\omega+\frac{i \vec{p}\cdot\vec{k}}{p^0}+\frac{1}{\tau_c}}\).
\end{split}
\ee

From Eq.\eqref{20}, we can show that $p_{k}\neq p_{l}$ and only $p_{x}$ and $p_{y}$ need to be considered. Subsequently,  we derive the dispersion relation in relativistic particle flow by substituting Eq. \eqref{19} and Eq. \eqref{20} into Eq. \eqref{6}, which yields
\be
\1\{\begin{split}
\label{21}
&\omega^2-k^2+2\kappa^2\int \frac{\dif^3p}{(2\pi)^3}\frac{(p_k p_l)^2 f'_0(p)}{p\epsilon_0}\(\frac{-i\omega}{-i\omega+\frac{i \vec{p}\cdot\vec{k}}{p^0}+\frac{1}{\tau_c}}\)=0,\\
&\omega^2-k^2-M^2+\frac{m^4\kappa^2}{6F(0)}\int \frac{\dif^3p}{(2\pi)^3}\frac{f'_0(p)}{p\epsilon_0}\( \frac{-i\omega}{-i\omega+\frac{i \vec{p}\cdot\vec{k}}{p^0}+\frac{1}{\tau_c}}\)=0.
\end{split}\2.
\ee

Furthermore, to determine mode's damping from dispersion, two damping mechanisms must be addressed:  Landau damping and collision-dominated hydrodynamic damping. These mechanisms are characterized by the imaginary part of the source. Landau damping refers to the excitation of two real particle-hole pairs caused by the decay of the mode, without considering the collision term. From the Eq. \eqref{19}, Eq. \eqref{20}, and the collisionless limit $\frac{1}{\tau_c}\rightarrow0$, we derive
\be
\1\{\begin{split}
\label{22}
&\Im{\(\frac{\Pi^{(1)}_{ij}}{\bar{h}^{TT}_{ij}}\)}=-\pi\omega\int \frac{\dif^3p}{(2\pi)^3}\frac{(p_k p_l)^2 f'_0(p)}{p\epsilon_0} \delta\(\omega-\frac{\vec{p}\cdot\vec{k}}{p^0}\),\\
&\Im{\(\frac{T^{(1)}}{\phi}\)}=\pi\omega m^4\int \frac{\dif^3p}{(2\pi)^3}\frac{f'_0(p)}{2p\epsilon_0} \delta\(\omega-\frac{\vec{p}\cdot\vec{k}}{p^0}\),
\end{split}\2.
\ee
The preceding formula shows that the Landau damping phenomenon happens only when $p^0=|p|\cos{\th}$ and the particles must be massless and move along the wave's direction to contribute. However, In flat spacetime, tensor mode ($\bar{h}^{TT}_{ij}$) will not encounter Landau damping because $(p_kp_l)^2=(p_xp_y)^2$.
Meanwhile, it is noteworthy to note that the Landau damping of the scalar mode ($\phi$) does not contribute too, as massless particles have $m=0$.

Additionally, to investigate another damping mechanism, we focus at the collision-dominated region ($\omega\ll\frac{1}{\tau_c}$), which follows
\be
\label{23}
\begin{split}
\Im{\Pi^{(1)}_{ij}}&=\bar{h}^{klTT}(\omega,\vec{k})\int \frac{\dif^3p}{(2\pi)^3}\frac{p_kp_lp_i p_j f'_0(p)}{2p\epsilon_0}\frac{-\frac{\omega}{\tau_c}}{\(\omega-\frac{\vec{p}\cdot\vec{k}}{p^0}\)^2+\frac{1}{\tau^2_c}}\\
&\approx-\frac{\omega\tau_c\bar{h}^{TT}_{ij}(\omega,\vec{k})}{15}\int\frac{\dif^3 p}{(2\pi)^3} \frac{p^3f'_0(p)}{\epsilon_0}=-\tau_c\omega\eta_1\bar{h}^{TT}_{ij}(\omega,\vec{k}),
\end{split}
\ee
and
\be
\label{24}
\begin{split}
\Im{T^{(1)}}&=m^4\phi(\omega,\vec{k})\int \frac{\dif^3p}{(2\pi)^3}\frac{f'_0(p)}{2p\epsilon_0}\frac{\frac{\omega}{\tau_c}}{\(\omega-\frac{\vec{p}\cdot\vec{k}}{p^0}\)^2+\frac{1}{\tau^2_c}}\\
&\approx m^4\omega\tau_c \phi(\omega,\vec{k})\int\frac{\dif^3p}{(2\pi)^3}\frac{f'_0(p)}{2p\epsilon_0}=\tau_c\omega\eta_2\phi(\omega,\vec{k}),
\end{split}
\ee
The collision-dominated viscosity coefficient under the relaxation time approximation are $\eta_1$ and $\eta_2$, which follows
\be
\1\{\begin{split}
\label{25}
&\eta_1\equiv\frac{\tau_c}{15}\int\frac{\dif^3 p}{(2\pi)^3} \frac{p^3f'_0(p)}{\epsilon_0},\\
&\eta_2\equiv m^4\tau_c\int\frac{\dif^3p}{(2\pi)^3}\frac{f'_0(p)}{2p\epsilon_0}.
\end{split}\2.
\ee
The viscosity coefficients are evidently based on the collision relaxation time and the distribution function of the equilibrium state. These two components are also the primary causes of the damping of tensor and scalar modes. Thus, we conclude that in flat spacetime, the evolution of $f(R)$ GW is predominantly influenced by collision damping, rather than by Landau damping.


\section{Damping of tensor and scalar modes in FRW universe}
In cosmology, the damping of tensor and scalar modes refer to the reduction in the amplitude of waves as the universe evolves. The energy of these waves changes as a result of cosmic expansion, which absorbs energy from the waves. Consequently, the frequency of waves is not constant. In flat spacetime, when the phase velocity of the wave deviates from the group velocity of excitations in matter, energy oscillates between the wave and the matter. However, precise cancellation occurs, with the energy transferred in one half-cycle of the wave being exactly counteracted by the loss in the other half-cycle, resulting in a net transfer rate of zero. Nevertheless, in an expanding universe, this cancellation is not entirely complete. This energy loss mechanism is distinct from Landau damping, which arises from processes such as photon diffusion, baryon acoustic oscillations, and gravitational interactions. 
Additionally, 
there is growing interest in exploring GW damping generated by alternative, non-inflationary sources, as proposed in other models. 
In this section, we will investigate damping of $f(R)$ GW within the FRW background.

We use conformal coordinates to investigate the evolution of tensor and scalar modes in FRW universe. The line element with a perturbed metric is represented by
\be
\label{26}
\dif s^2=a^2(\tau)\[-\dif \tau^2+\(\delta_{ij}+\bar{h}^{TT}_{ij}\)\dif x^i\dif x^j\],
\ee
 and the cosmological equation satisfied by the tensor mode could be determined by \cite{Odintsov:2021kup,DeFelice:2010aj}
\be
\label{27}
\ddot{\bar{h}}^{TT}_{ij}+\(2+a_M\)H(\tau)\dot{\bar{h}}^{TT}_{ij}-\na^2\bar{h}^{TT}_{ij}=2\kappa^2a^2(\tau)\Pi^{(1)}_{ij}.
\ee
From Eq. \eqref{6}, the scalar mode in the $f(R)$ model can also be expressed as

\be
\label{28}
\ddot{\phi}+\(2H(\tau)+\frac{2\dot{F}}{F}\)\dot{\phi}-\na^2\phi+\(a^2(\tau)M^2+2H(\tau)\frac{\dot{F}}{F}+\frac{\ddot{F}}{F} \)\phi=-\frac{\kappa^2}{3F(R_0)}a^2(\tau)T^{(1)}.
\ee
Where $M^2\equiv \frac{1}{3}\(\frac{F(R_0)}{F'(R_0)} -R_0\)$, $\dot{\bar{h}}^{TT}_{ij}$ and $\dot{\phi}$ denote the derivatives with respect to the conformal time $\tau$, and $H$ indicates the Hubble constant. $a_M$ is defined as $\frac{F'(R_0)\dot{R}}{F(R_0)H}$, with the scenario of $a_M=\phi=0$ corresponding to GR. We focus on the conclusion derived from the right-hand side of the above equation. The perturbation of the Boltzmann equation \eqref{15} is
\be
\label{29}
\frac{1}{a(\tau)}\frac{\p\delta f}{\p\tau}+\frac{p^m\p_m\delta f}{a(\tau)p^{\tau}}+\frac{1}{a(\tau)}\frac{\dif p_m}{\dif\tau}\frac{\p f_0}{\p p_m }=-\frac{1}{\tau_c}\(\delta f-\delta f_h\),
\ee
which can be simplify to
\be
\label{30}
\(\frac{\p}{\p\tau}+v^m\p_m+\frac{1}{\bar{\tau}_c}\)\delta f=\frac{\delta f_h}{\bar{\tau}_c}-\frac{\dif p_m}{\dif\tau}\frac{\p f_0}{\p p_m },
\ee
where $v^m\equiv\frac{ p^{m}}{ p^{\tau}}=\frac{p_m}{\sqrt{p^2+m^2a^2}}$ corresponds to the three-velocity of particles. $\bar{\tau}_c\equiv\frac{\tau_c}{a(\tau)}$ is collision time in cosmology. The term $\frac{\dif p_m}{\dif\tau}$ is given by
\be
\label{31}
\begin{split}
\frac{\dif p_m}{\dif\tau}&=\frac{1}{2}\p_m g_{\m\n}\frac{p^{\m}p^{\n}}{p^{\tau}}\\
&=\frac{\p_m a^4(\tau)\bar{h}^{ijTT}p_ip_j+m^2a^2(\tau)\p_m\phi}{2p^{\tau}a^2(\tau)}.
\end{split}
\ee
Similarly, the on-shell condition and its perturbation could be denoted by
\be
\1\{\begin{split}
\label{32}
&\epsilon=\epsilon_0+\delta\epsilon,\\
&\epsilon_0=\sqrt{m^2a^2(\tau)+p^2},\\
&\delta\epsilon=\frac{a^2(\tau)\delta g^{\m\n}p_{\m}p_{\n}}{2\epsilon_0}=\frac{-a^4(\tau)\bar{h}^{ijTT}p_ip_j-m^2a^2(\tau)\phi}{2\epsilon_0}.
\end{split}\2.
\ee
Where $\delta g^{\m\n}\equiv\frac{h^{\m\n}}{a^2}$, and the spatial Fourier transform is used to simplify the final Boltzmann equation \eqref{30} to
\be
\label{33}
\(\frac{\p}{\p\tau}+Q(\tau)\)\delta f(\tau,\vec{k})=-\frac{f'_0(p)Q(\tau)}{2p}\[a^4(\tau)\bar{h}^{ijTT}(\tau,\vec{k})p_ip_j+m^2a^2(\tau)\phi(\tau,\vec{k})\],
\ee
with
\be
\label{34}
Q(\tau)\equiv i\vec{v}\cdot\vec{k}+\frac{1}{\bar{\tau}_c}.
\ee
Then, the particular solution of the first-order differential equation \eqref{33} can be written as
\be
\1\{\begin{split}
\label{35}
&\delta f(\tau)=-\int^{\tau}_{\tau_0}\(a^4(\tau')\bar{h}^{ijTT}(\tau',\vec{k})p_ip_j+m^2a^2(\tau')\phi(\tau',\vec{k})   \)\frac{f'_0(p)}{2p}\frac{\p e^{-\Lambda\(\tau,\tau'\)}}{\p\tau'}\dif\tau',\\
&\Lambda\(\tau,\tau'\)\equiv\Lambda_1\(\tau,\tau'\)+i\cos{\th}k\Lambda_2\(\tau,\tau'\)=\int^{\tau}_{\tau'}\frac{1}{\bar{\tau}_c(\tau'')}\dif \tau''+i\cos{\th}k\int^{\tau}_{\tau'}v(\tau'')\dif \tau'' ,\\
\end{split}\2.
\ee
where $\tau_0$ depicts the initial assertion at which the system is in equilibrium, and the distribution function
\be
\label{36}
f_{0}=\frac{g}{e^{\frac{p^t}{\mathscr{T}}}\pm 1}=\frac{g}{e^{\frac{\epsilon_{0}}{a_0\mathscr{T}_0}}\pm 1},
\ee
where the scale factor is set to zero ($a_0=1$) in the present universe. $\mathscr{T}_0$ represents the current background radiation temperature. The perturbed anisotropic part is a generalization of Eq. \eqref{17}, which is \cite{Flauger:2017ged}
\be
\1\{\begin{split}
\label{37}
&\Pi^{(1)}_{ij}=\int \frac{\dif^3p}{(2\pi)^3}\frac{p_ip_j}{\sqrt{-g}\epsilon_0}\bar{\delta} f=\int \frac{\dif^3p}{(2\pi)^3}\frac{p_ip_j}{a^4\sqrt{m^2a^2+p^2}}\bar{\delta} f,\\
&T^{(1)}=-m^2\int \frac{\dif^3p}{(2\pi)^3}\frac{1}{\sqrt{-g}\epsilon_0}\bar{\delta}f=-m^2\int \frac{\dif^3p}{(2\pi)^3}\frac{1}{a^4\sqrt{m^2a^2+p^2}}\bar{\delta} f ,
\end{split}\2.
\ee
with
\be
\label{38}
\bar{\delta} f=\int^{\tau}_{\tau_0}e^{-\Lambda(\tau,\tau')}\frac{\p}{\p\tau'}\[\frac{f'_0(p)}{2p}\(a^4(\tau')\bar{h}^{ijTT}(\tau',\vec{k})p_ip_j+m^2a^2(\tau')\phi(\tau',\vec{k})\)\]\dif\tau' .
\ee
Substituting Eq. \eqref{38} into Eq. \eqref{37} and utilizing the following integration formula  \cite{Flauger:2017ged,Stefanek:2012hj}
\be
\1\{\begin{split}
\label{39}
&\int^{2\pi}_{0}\dif\phi \frac{p_ip_jp_kp_l}{p^4}=\frac{\pi(1-\cos^2{\th})^2}{4}\(\delta_{ij}\delta_{kl}+\delta_{ik}\delta_{jl}+\delta_{il}\delta_{jk}\),\\
&K(x)\equiv\frac{1}{16}\int^{1}_{-1}\(1-\cos^2{\th}\)^2e^{ix\cos{\th}}\dif \cos{\th},
\end{split}\2.
\ee
we obtain
\be
\1\{\begin{split}
\label{40}
&\Pi^{(1)}_{ij}=\int\frac{p^5\dif p}{2\pi^2a^4\sqrt{m^2a^2+p^2}}\int^{\tau}_{\tau_0}K\(k\Lambda_2(\tau,\tau')\)e^{-\Lambda_1(\tau,\tau')}\frac{\p}{\p\tau'}\[f'_0(p)\bar{h}^{TT}_{ij}(\tau')\]\dif\tau',\\
&T^{(1)}=-m^4\int \frac{p\dif p}{4\pi^2a^4\sqrt{m^2a^2+p^2}}\int^{\tau}_{\tau_0}\frac{\sin{k\Lambda_2(\tau,\tau')}}{k\Lambda_2(\tau,\tau')}e^{-\Lambda_1(\tau,\tau')}\frac{\p}{\p\tau'}\[f'_0(p)a^2(\tau')\phi(\tau')\]\dif\tau'.
\end{split}\2.
\ee
Where $K(x)\equiv-\frac{\sin{x}}{x^3}-\frac{3\cos{x}}{x^4}+\frac{3\sin{x}}{x^5}=\frac{j_0(x)}{15}+\frac{2j_{2}(x)}{21}+\frac{j_{4}(x)}{35}$ is a linear combination of spherical Bessel functions \cite{Stefanek:2012hj}. From the above equation, it is observed that the scalar and tensor modes in the $f(R)$ theory do not couple together. They independently influence their own evolution equations, consistent with flat spacetime. When $m=0$, the tensor mode's equation returns to the Weinberg's conclusion, although with additional collision contribution. However, it is noteworthy that the contribution of the additional scalar mode and their evolution merits attention. Since the existence of Eq. \eqref{40}, the mode involves the Landau and Collision damping phenomena.

\section{Numerical solution of damping from neutrinos }
In the previous section, we established the damping equations for $f(R)$ GW. In this section, we will investigate the damping of waves within the neutrino system, focusing on the effects of mass.

Neutrinos are fundamental Fermi particles that participated weak and gravitational interactions during the early stage of the Big Bang. Before decoupling, the interactions of neutrinos reached chemical equilibrium, leading neutrinos to follow an equilibrium state distribution function. Initially, Weinberg's original research focused on the effect of three massless neutrinos. Subsequently, recent cosmological development have suggested deviations from the traditional assumption of three effective neutrino degrees of freedom. Experimental evidence from neutrino oscillations confirms that neutrinos have mass, which could influence gravitational wave's damping. Therefore, we intend to explore the impact of neutrinos mass on the evolution of the modes.

Additionally, the damping phenomenon occurs when the $k$ of two modes are longer than the cosmic horizon $k_{eq}\equiv a(eq)H(eq)$ ($\tau_{eq}$ represents the time when the proportion of radiation and matter are the same). During the radiation and matter dominated period, the energy density of the neutrinos are still mostly manifested as $a^{-4}$. Hence, with the increase of the scale factor $a(\tau)$, the damping effect induced by $\Pi_{ij}$ diminishes gradually. Meanwhile, as indicated by Eq. \eqref{40}, the emergence of collision term also gradually eliminates the contribution of the anisotropic tensor.

Now, we focus on the evolution of two modes with $f(R)=R+\al R^2$. Based on the geodetic precession measured by the Gravity Probe B experiment, the parameter has been constrained to $\alpha < 5 \times 10^{11}\text{m}^2$, whereas for the pulsar B in the PSR J0737-3039 system the bound is about $10^4$ times larger \cite{PhysRevD.81.104003}. Through the research of planetary precession rates, the parameter has been constrained to $\alpha < 0.6 \times 10^{18} \, \text{m}^2$ \cite{Berry:2011pb}. The upper limit of the graviton mass given by the LIGO observation is $M <1.2 \times 10^{-22} \, \text{eV}$ $\(M\equiv\frac{1}{6\alpha}\)$, and a more stringent limit from the dynamics of the galaxy cluster is $M < 2 \times 10^{-29} \, \text{eV}$ \cite{PhysRevD.9.1119}. Furthermore, the constraint on the mass from the new solution of the ephemeris INPOP19a is $M < 3.16 \times 10^{-23} \, \text{eV}$ at the $90\%$ confidence level \cite{Gao:2022hho}. We will expore the evolution of GW within the parameter range $0 < \al < 10^{18} \, \text{m}^2$.  

After Fourier transforming the spatial part, Eqs. \eqref{27} and \eqref{28} are transformed into
\be
\1\{\begin{split}
\label{41}
&\ddot{\bar{h}}^{TT}_{ij}(u)+2H(u)\dot{\bar{h}}^{TT}_{ij}(u)+\frac{2\al\dot{R}}{1+2\al R}\dot{\bar{h}}^{TT}_{ij}(u)+\bar{h}^{TT}_{ij}(u)=\frac{2\kappa^2T^4_0}{k^2}a^2(u)\Pi^{(1)}_{ij},\\
&\Pi^{(1)}_{ij}=\int^{\infty}_{0}\frac{x^5\dif x}{2\pi^2a^4\sqrt{\frac{m^2a^2(u)}{T^2_0}+x^2}}\int^{u}_{u_{dec}}K\(\Lambda_2(u,u')\)e^{-\Lambda_1(u,u')}\frac{\p}{\p u'}\[ \frac{\dif f_0(x,u')}{\dif x}\bar{h}^{TT}_{ij}(u')\]\dif u',
\end{split}\2.
\ee
and
\be
\1\{\begin{split}
\label{42}
&\ddot{\phi}(u)+\(2H(u)+\frac{4\al\dot{R}}{1+2\al R}\)\dot{\phi}(u)+\(1+\frac{a^2(u)}{6\al k^2}+\frac{4\al H(u)\dot{R}}{1+2\al R}+\frac{2\al\ddot{R}}{1+2\al R}\)\phi(u)\\
&=-\frac{\kappa^2a^2(u)}{3k^2(1+2\al R)}T^{(1)},\\
&T^{(1)}=-m^4\int^{\infty}_{0}\frac{x\dif x}{4\pi^2a^4\sqrt{\frac{m^2a^2}{T^2_0}+x^2}}\int^{u}_{u_{dec}}\frac{\sin{\Lambda_2(u,u')}}{\Lambda_2(u,u')}e^{-\Lambda_1(u,u')}\frac{\p}{\p u'}\[\frac{\dif f_0(x,u')}{\dif x}a^2(u')\phi(u')\]\dif u',
\end{split}\2.
\ee
with
\be
\1\{\begin{split}
\label{43}
&\Lambda_1\(u,u'\)=\int^{u}_{u'}\frac{1}{k\bar{\tau}_c(u'')}\dif u'',\\
&\Lambda_2\(u,u'\)=\int^{u}_{u'}v(u'')\dif u'',
\end{split}\2.
\ee
where we introduce the dimensionless independent variables $u \equiv k\tau$, $x\equiv\frac{p}{T_0}$, and neutrinos decoupling time $u_{dec}$. To visually illustrate the impact of incorporating nonzero neutrino masses, we employ the straightforward analytical expression for the scale factor in a universe dominated by matter and radiation, which provided by
\be
\label{44}
a(u)=\frac{u^2}{\bar{u}^2}+2\frac{u}{\bar{u}}\sqrt{a_{eq}},
\ee
with
\be
\label{45}
\bar{u}\equiv \frac{2k}{\sqrt{\Omega_M}H_0}.
\ee
In standard cosmic evolution, there are three generations of neutrinos corresponding to $a_{eq}=\frac{1}{3600}$, $\Omega_M=0.3$ \cite{PhysRevD.73.123515,Dent:2013asa}.
According to the above equation, Eq. \eqref{43} can be stated as
\be
\1\{\begin{split}
\label{46}
&\Lambda_1\(u,u'\)=\frac{u^3-u'^3}{3\bar{u}^2k\tau_c}+\frac{(u^2-u'^2)\sqrt{a_{eq}}}{\bar{u}k\tau_c},\\
&\Lambda_2\(u,u'\)=\int^{u}_{u'}\frac{x}{\sqrt{x^2+\frac{m^2a^2(u'')}{T^2_0}}}\dif u''.
\end{split}\2.
\ee

According to Eq. \eqref{41} and Eq. \eqref{42}, it can be deduced that a nonzero neutrino will also exert a certain influence on the two modes as their wavelengths enter the cosmological horizon. The Fig.1 depicts the numerical results of the tensor and scalar modes. We emulate Weinberg's initial approach by defining $\chi(u)\equiv\frac{\bar{h}^{TT}_{ij}(u)}{\bar{h}^{TT}_{ij}(0)}$ $\(\text{or} \frac{\phi(u)}{\phi(0)}\)$  to represent the vertical axis and dimensionless evolutionary time $u\equiv k\tau$ to represent the horizontal axis. Particularly, the neutrino decoupling time has been set as the initial moment for $\chi(u)$. The first four figures illustrate the evolution of tensor mode, with contributions considered from both cases: no matter and neutrino masses with $0$ and $1 \, \text{eV}$.  The final two figures depict the evolution of scalar mode.

\begin{figure}[H]
\centering
\begin{minipage}{0.5\textwidth}
\centering
\includegraphics[scale=0.6,angle=0]{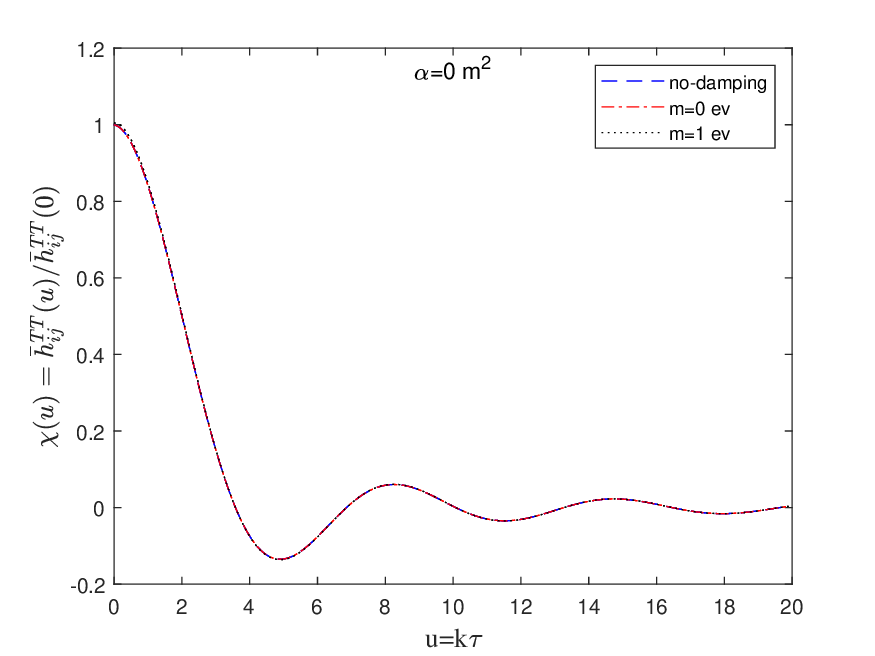}
\end{minipage}%
\centering
\begin{minipage}{0.5\textwidth}
\centering
\includegraphics[scale=0.6,angle=0]{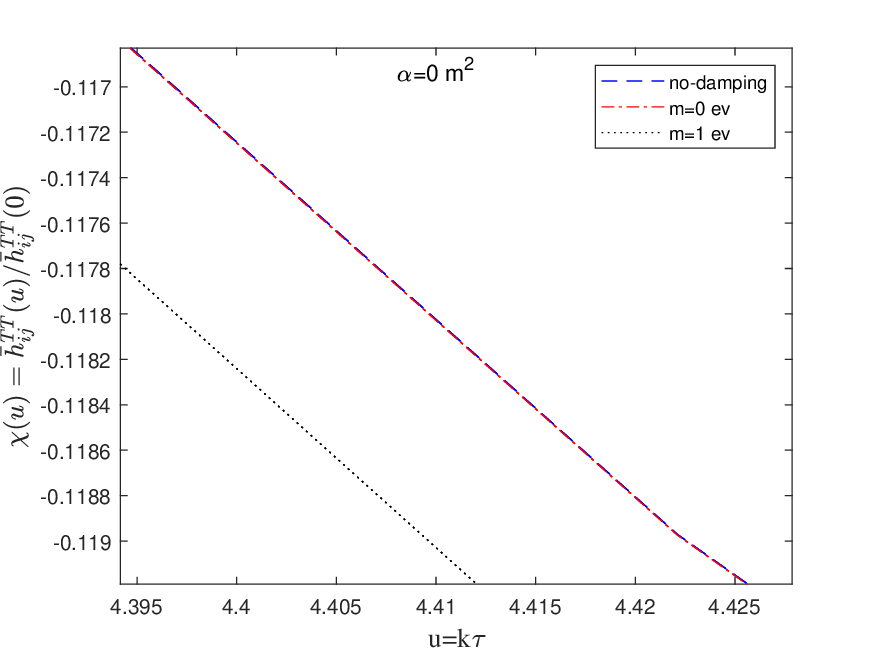}
\end{minipage}
\begin{minipage}{0.5\textwidth}
\centering
\includegraphics[scale=0.6,angle=0]{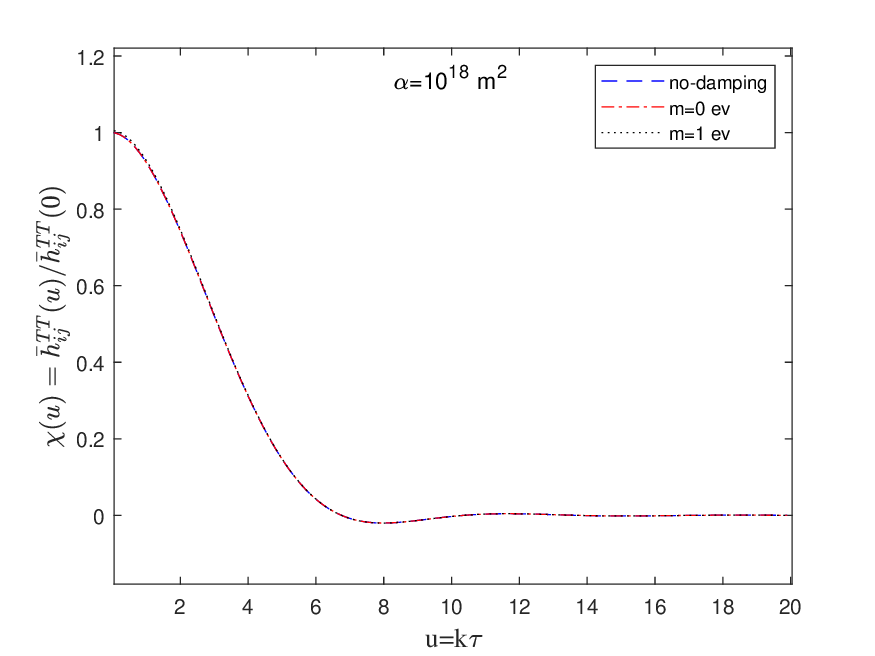}
\end{minipage}%
\begin{minipage}{0.5\textwidth}
\centering
\includegraphics[scale=0.6,angle=0]{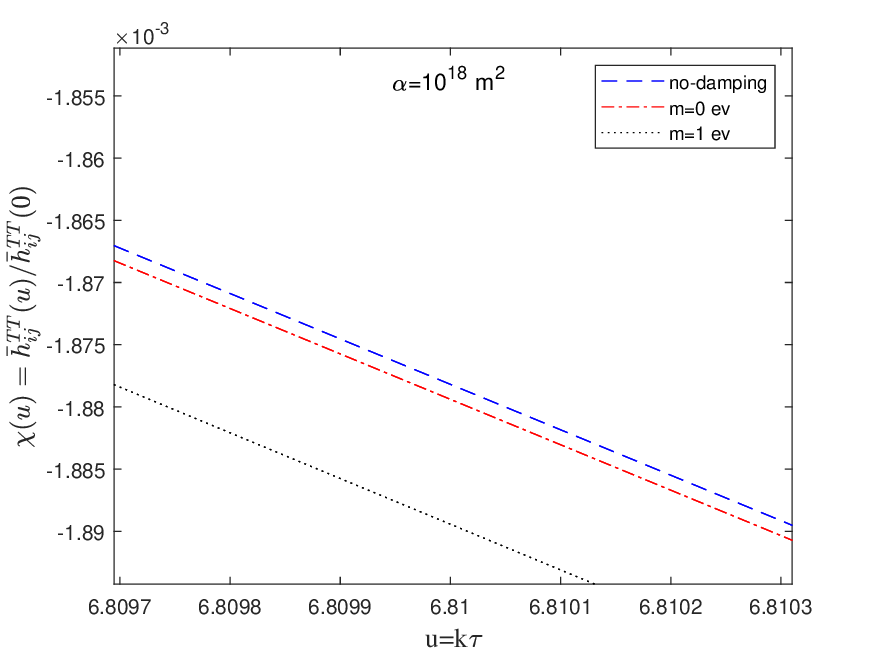}
\end{minipage}
\begin{minipage}{0.5\textwidth}
\centering
\includegraphics[scale=0.6,angle=0]{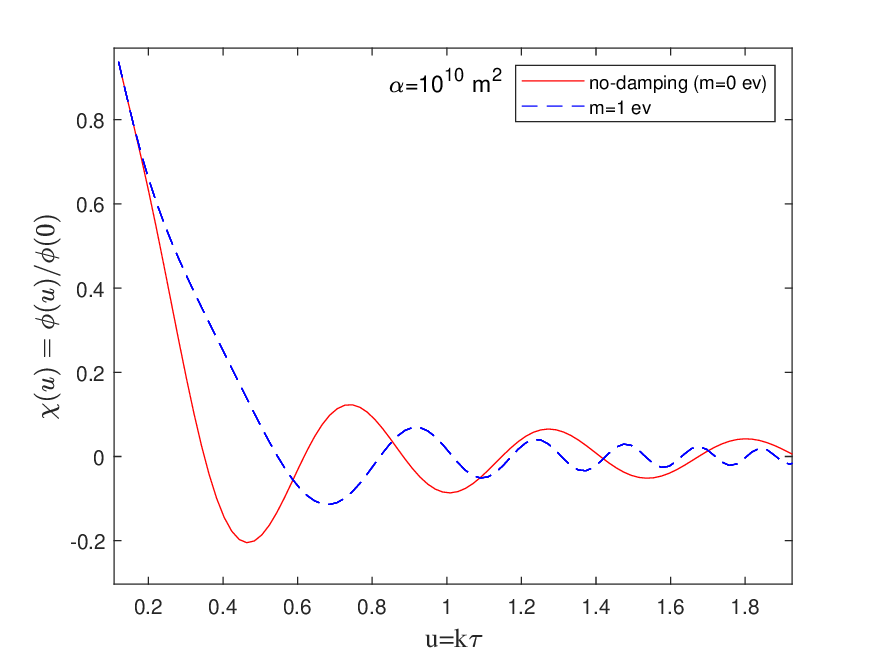}
\end{minipage}%
\begin{minipage}{0.5\textwidth}
\centering
\includegraphics[scale=0.6,angle=0]{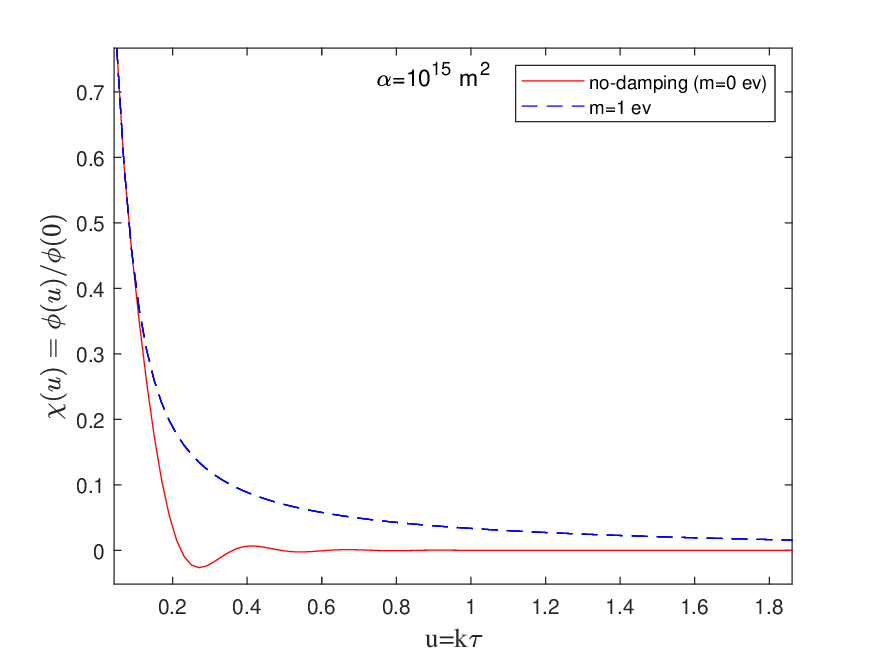}
\end{minipage}
\caption{\label{fig.2} The top-left and bottom-left pictures mainly describe the evolution of the tensor mode. The top-right and bottom-right plots depict enlarged versions of the figures. The last two figures demonstrate the impact of neutrino mass on the evolution of scalar mode with $k\tau_c=100$, $\bar{u}=100$. }
\end{figure}

The analysis of the first four figures reveal that the impact of neutrino mass on damping is subtle.
 Closer inspection shows that the case with $m = 1$ eV exhibits a slightly more rapid damping rate compared to $m = 0$ eV. This suggests that neutrinos with $m = 1$ eV introduce a subtle damping effect on the waves. However, the difference is on the order of  $10^{-3} \sim 10^{-4}$, indicating that detecting such mass-induced variations will be challenging. In contrast, the differences in the scalar mode shown in the last two figures are more pronounced. The corresponding damping due to mass directly inhibits wave attenuation, resulting in a slower oscillation frequency, with this attenuation occurring over a very short timescale. Furthermore, the parameter $\al$ directly reduces the wave amplitude. Specifically, when $\alpha < 10^{15} \text{ m}^2$, oscillatory damping patterns begin to emerge.

\section{conclusion and discussion}
In this work, we applied kinetic theory to investigate the damping behavior of $f(R)$ GW in the presence of medium. Firstly, we introduced the linearized $f(R)$ model and constructed wave equation for the scalar mode. Subsequently, we calculated the perturbed form of the Boltzmann equation, obtained the solution in momentum space, and  incorporated it into the transverse-traceless part of the anisotropic stress tensor $\Pi_{ij}$ (or the trace of energy-momentum tensor $T$) to establish the dispersion relation. Additionally, we examined the damping coefficient in the collision-dominated regime and Landau damping in the collisionless limit. Our findings revealed  that the Landau damping contributions from both tensor and scalar modes were zero.

Subsequently, we examined the Boltzmann equation governing the perturbations in the FRW scenario and derived the wave equations for the tensor and scalar modes, including their damping effects. Moreover, after the decoupling of neutrinos, we numerically solved the decay of GW. For $f(R)=R+\al R^2$, we explore how the mass term influences the decay of wave amplitude in the neutrino system. Our findings indicate that the tensor mode with $m = 1 \, \text{eV}$ decays more rapidly than in other scenarios, whereas the scalar mode with $m = 1 \, \text{eV}$ seems to suppress decay. 



\begin{acknowledgments}
This work was supported by the National Key R\&D Program of China (Grants No. 2022YFA1403700),  NSFC (Grants No. 12141402, 12334002, 12333008), the SUSTech-NUS Joint Research Program, Center for Computational Science and Engineering at Southern University of Science and Technology, and Hebei Provincial Natural Science Foundation of China (Grant No. A2021201034).
\end{acknowledgments}

\appendix
\bibliographystyle{unsrt}
\bibliography{damping}

\end{document}